\documentstyle[12pt,moriond,psfig]{article}
\begin{document}
\heading{Big Bang nucleosynthesis and the baryonic content of the universe}

\author{Trinh Xuan Thuan$^{1}$ \& Yuri I. Izotov $^{2}$} 
{$^{1}$ Astronomy Department, University of Virginia, Charlottesville, VA 22903
, USA}  {$^{2}$ Main Astronomical Observatory, Golosiiv, Kyiv 03680, Ukraine}

\begin{moriondabstract}

A review of the latest measurements of the primordial abundances of the 
light nuclei D, $^3$He, $^4$He and $^7$Li is given. We discuss in 
particular the primordial abundance $Y_{\rm p}$ of $^4$He as measured
in blue compact dwarf galaxies. We argue that the best measurements 
now give a ``high'' value of $Y_{\rm p}$ along with a ``low'' value of D/H,
and that the two independent measurements are consistent within the framework 
of standard Big Bang nucleosynthesis with a number of light neutrino 
species $N_\nu$ = 3.0$\pm$0.3 (2$\sigma$).

\end{moriondabstract}

\section{Introduction}

The Big Bang theory which says that the universe began its existence from 
an inimaginably small, hot and dense state is supported by four main 
key observations: 1) the expansion of the universe, 2) the Planck blackbody 
spectrum of the Cosmic Microwave Background (CMB), 3) the tiny density 
fluctuations in the CMB and the resulting large-scale distribution of galaxies, and 
4) the chemical make-up of the stars and galaxies, with which we will be 
concerned here.
In the Standard Big Bang Nucleosynthesis (SBBN) model, light nuclei H, D, 
$^3$He, $^4$He, and $^7$Li were produced by nuclear reactions a few 
minutes after the birth of the universe. Given the number of light neutrino 
species $N_\nu$ = 3 and the neutron lifetime, 
the abundances of
these light elements depend on one cosmological parameter only, the
baryon-to-photon ratio $\eta$, which in turn is directly related to the
density of ordinary baryonic matter $\Omega_{\rm b}$. The ratio of any two 
primordial abundances, for example that of $^4$He to H gives $\eta$, 
and accurate measurements of the other three light elements tests SSBN.  

$\eta$ is determined during baryogenesis which occurs when three conditions 
are fulfilled in the early universe, as pointed out by Sakharov (1967): 
1) a symmetry violation leading to 
different interactions for matter and anti-matter (such as CP violation), 
2) interactions which modify the baryon number and 3) departure from 
thermodynamic equilibrium. It is not known when the last condition occurs, but
if it occurs in a first order phase transition, at the electroweak scale,
then it may be hoped that future experiments will allow to predict $\eta$.
On the other hand, if it occurs at the GUT or inflation scale, which is 
not accessible to experimentation in the near-future, predicting $\eta$ will 
be very difficult.  
 
The main physical processes in SBBN are the following (Kolb \& Turner 1990,
Tytler et al. 2000):
 at early times, for times less than about 1 second, 
weak reactions maintain the n/p ratio close to the Boltzman value. As the 
universe expands and the temperature drops, n/p decreases until about 1 
second ($T$ $\sim$ 1MeV), when the weak reaction rate becomes slower than the 
expansion rate. At that time, the n/p ratio freezes at the value of about 1/6.
n and p start to combine to make D, but photodissociation is rapid and no 
significant build-up of light nuclei occurs until about 100 seconds, when 
the universe has cooled down to about 0.1 MeV, well below the binding 
energies of the light nuclei. About 20\% of the free neutrons have decayed 
before being included in nuclei. The remaining neutrons go into the building 
of $^4$He nuclei. The primordial helium mass fraction $Y_{\rm p}$ of $^4$He 
depends relatively weakly on $\eta$ as the n/p ratio depends on weak reactions 
between nucleons and leptons and not on pairs of nucleons. If $\eta$ is 
larger, $Y_{\rm p}$ increases because nucleosynthesis starts earlier and more 
nucleons end up in $^4$He nuclei, and less in D and $^3$He. As for 
$^7$Li, two channels contribute to manufacture it in the $\eta$ range of 
interest, so that a given $^7$Li abundance corresponds to two different values 
of $\eta$.  

Section 2 discusses our latest measurements of the primordial $^4$He 
abundance as derived from observations of
 blue compact dwarf galaxies. Section 3 reviews the 
measurements of the primordial abundances of other light elements. 
Section 4 discusses whether these independent 
measurements are consistent with each other within 
the framework of SSBN, and gives the derived cosmological baryon density.
Section 5 discusses the number of light neutrino species as derived from 
the most consistent set of primordial element abundances.

\section{The primordial $^4$He abundance as derived from blue compact dwarf 
galaxies}

Blue compact dwarf galaxies (BCD) are low-luminosity ($M_B$ $\geq$ --18) 
systems which are undergoing an intense burst of star formation in a very 
compact region (less than 1 kpc) which dominates the light of the galaxy 
(Figure 1) and which shows blue colors and a HII region-like emission-line 
optical spectrum (Figure 2). BCDs are ideal laboratories in which to measure 
the primordial $^4$Helium abundance because of several reasons:

1) With an oxygen abundance O/H ranging between 1/50 and 1/3 that of the 
Sun, BCDs are among the most metal-deficient gas-rich galaxies known.
Their gas has not been processed through many generations of stars, and thus   
best approximates the pristine primordial gas. Izotov \& Thuan (1999) have 
argued that BCDs with O/H less than $\sim$ 1/20 that of the Sun may be genuine 
young galaxies, with stars not older than $\sim$ 100 Myr. Thus the primordial 
Helium mass fraction $Y_{\rm p}$ can be derived accurately in very 
metal-deficient BCDs with only a small correction for Helium made in stars.

%
%
\def\baselinestretch{0.5}
\begin{figure}[tbh]
\centering
\hspace*{0.5cm}\psfig{figure=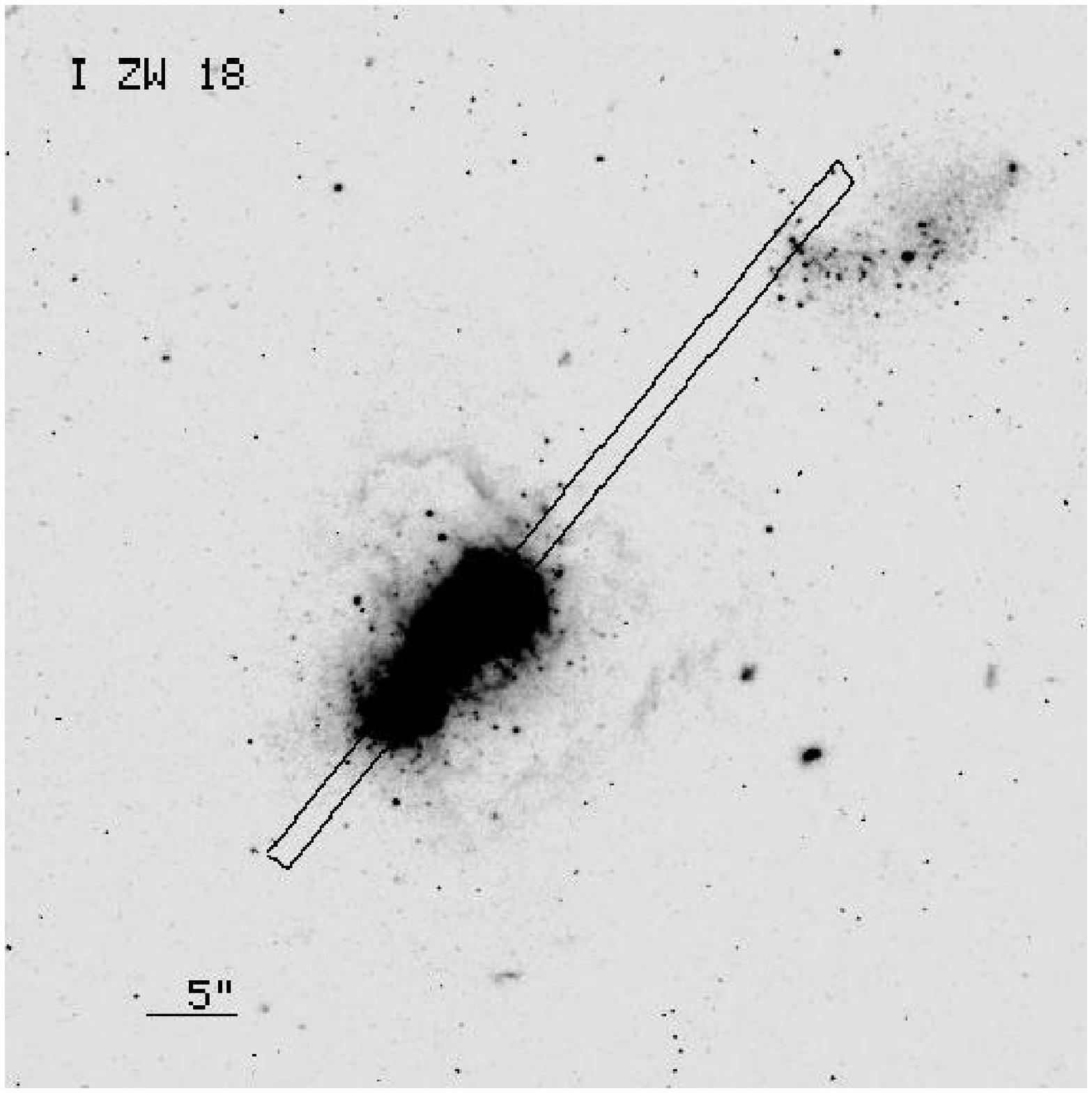,width=5.0cm,angle=0}
\hspace*{0.5cm}\psfig{figure=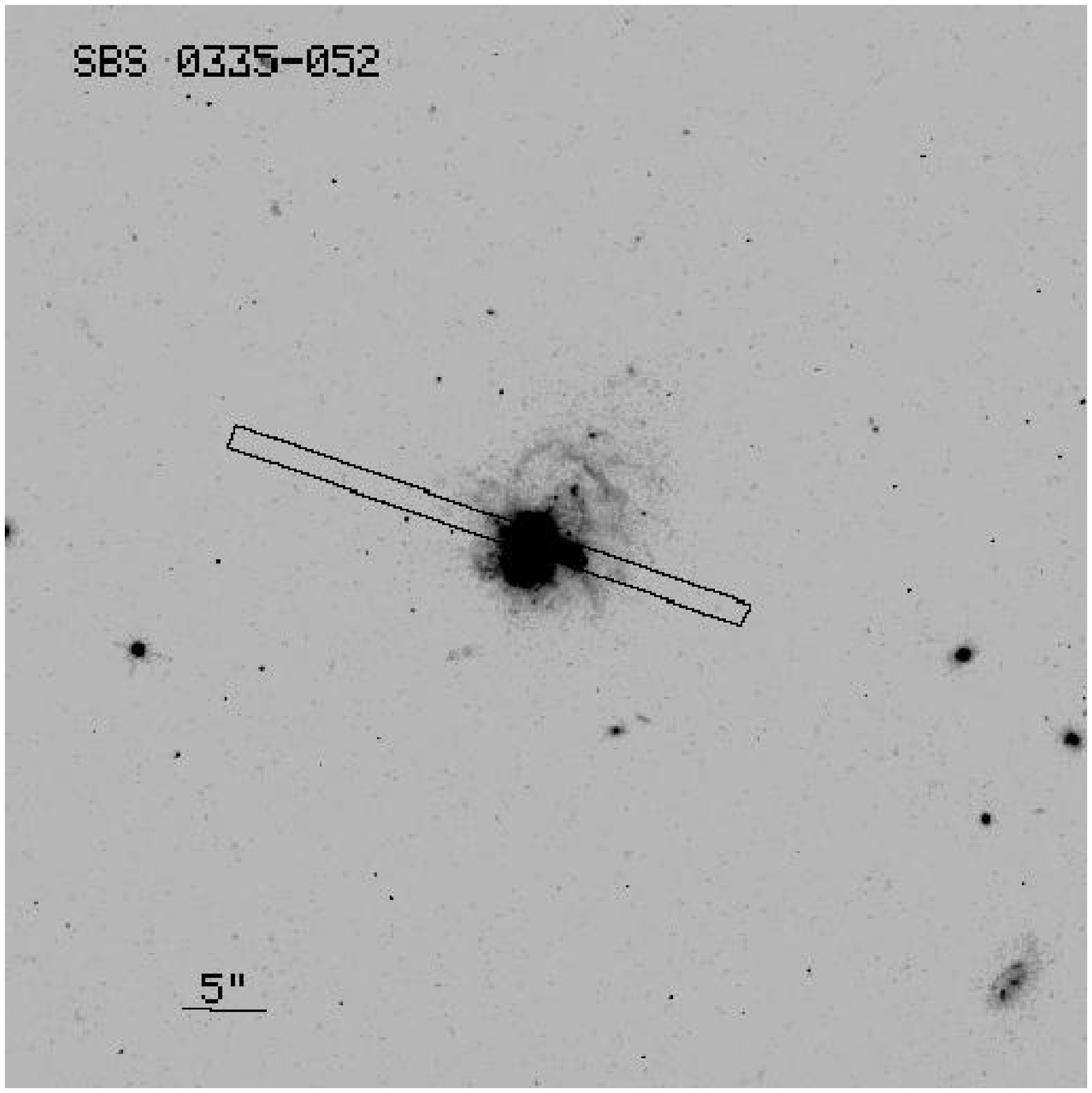,width=5.0cm,angle=0}
\caption[]{\label{Fig1} Hubble Space Telescope  $V$ images of 
of the two most metal-deficient blue compact dwarf galaxies known: 
I Zw 18 (1/50 solar) and SBS 0335--052 (1/41 solar).
The spatial scale is 1 arcsec = 49 pc in the case of I Zw 18 and is
1 arcsec = 257 pc in the case of SBS 0335--052.  }
\end{figure}
\def\baselinestretch{1.0}

%

\def\baselinestretch{0.5}
\begin{figure}[tbh]
\centering
\hspace*{-0.5cm}\psfig{figure=thuan_fig2.ps,width=7.0cm,angle=0}
\hspace*{0.5cm}\psfig{figure=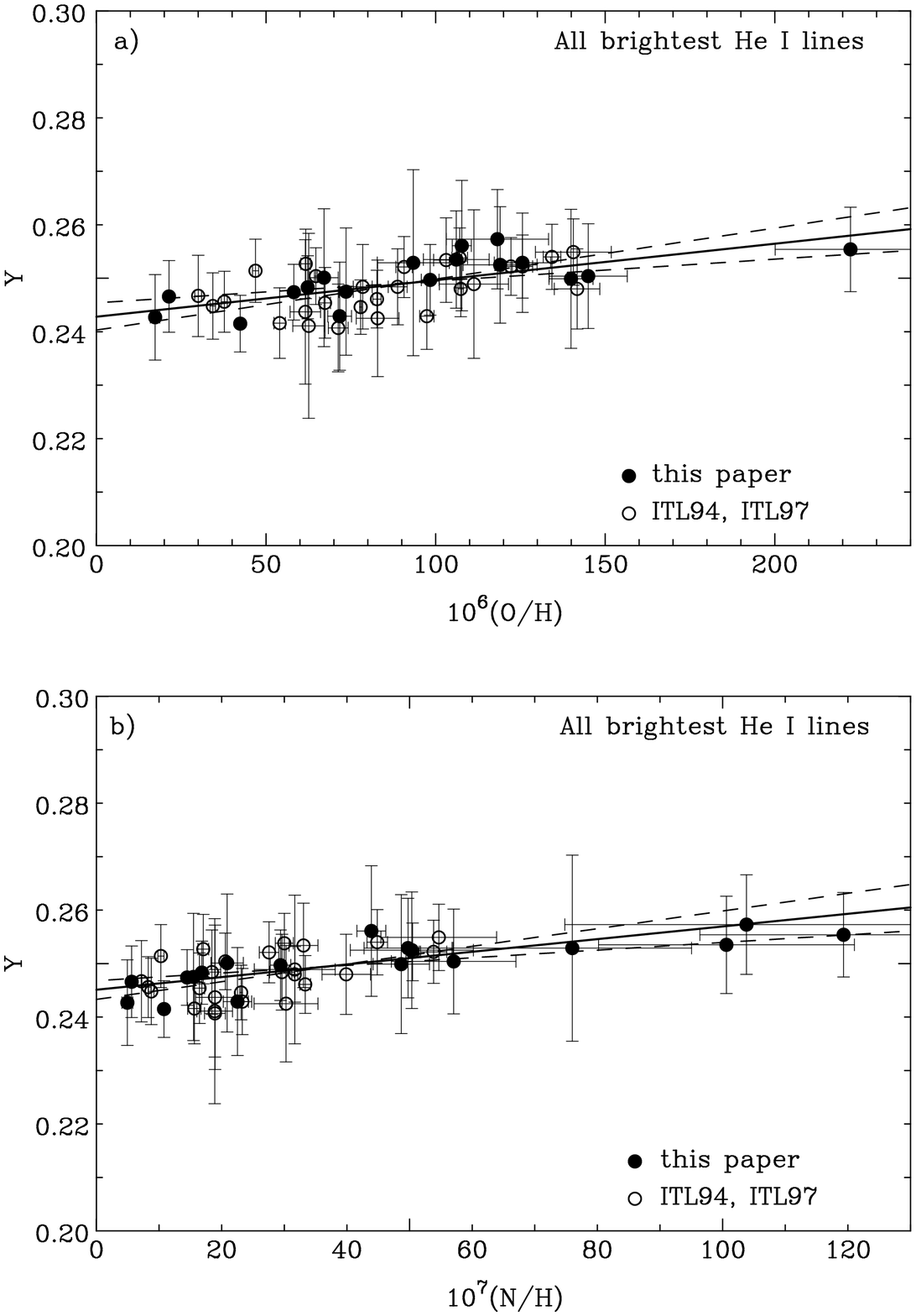,width=7.0cm,angle=0}
\caption[]{\label{Fig2} Spectra
of the brightest parts of the NW and the SE components of I Zw 18. 
 The positions of He I lines used for $^4$He abundance determination
are marked. Note the effect of underlying stellar absorption is 
much larger in the NW than in the SE component: 
all marked He I lines in the spectrum of the SE
component are in emission while the two He I $\lambda$4026 and $\lambda$4921
lines are in absorption and the He I $\lambda$4471 emission line is 
barely detected in the spectrum of the NW component. }
\caption[]{\label{Fig3}Linear regressions of (a) the helium mass fraction
$Y$ vs. oxygen abundance O/H and (b) the helium mass fraction $Y$ vs.
nitrogen abundance for our sample of
45 H II regions. The $Y$s are derived self-consistently 
by using the 5 brightest He I emission lines in the optical range.
Collisional and fluorescent enhancements, underlying He I stellar absorption and
Galactic Na I interstellar absorption are taken into account. }
\end{figure}

\def\baselinestretch{1.0}

2) Because of the relative insensitivity of $^4$He production to the 
baryonic density of matter,  $Y_{\rm p}$ needs to be 
determined to a precision better than 5\% to provide useful cosmological 
constraints. This precision can in principle be achieved by using BCDs because
their optical spectra show several He I recombination emission lines and  
very high signal-to-noise ratio emission-line spectra with moderate spectral 
resolution of BCDs can be obtained at large telescospes (4 m class or larger) 
coupled with efficient and linear CCD detectors with a relatively modest 
investment of telescope time. The theory of nebular emission is well 
understood and the theoretical He I 
recombination coefficients are well known enough to allow to 
convert He emission-line strengths into abundances with the desired accuracy.
 
 $Y_{\rm p}$ is generally determined  by 
linear extrapolation of the correlations $Y$--O/H and $Y$--N/H to O/H = N/H = 0 
(Peimbert \& Torres-Peimbert 1974,  Pagel, Terlevich 
\& Melnick 1986), where $Y$, N/H and O/H are respectively the 
$^4$He mass fraction, the Oxygen and Nitrogen abundances relative to Hydrogen
of a sample of dwarf irregular and BCD galaxies.
Izotov, Thuan \& Lipovetsky (1994, 1997), Izotov \& Thuan (1998) and 
Izotov et al. (1999) have obtained high 
signal-to-noise ratio spectra for a relatively large sample of $\sim$ 45 BCDs
(see Thuan \& Izotov 2000 for a review).
  
 We obtain $Y_{\rm p}$= 0.2443$\pm$0.0015 with d$Y$/d$Z$ = 2.4$\pm$1.0 
(Figure 3).  Our $Y_{\rm p}$ 
is considerably higher than those derived in previous work 
by other groups which range 
from 0.228$\pm$0.005 (Pagel et al. 1992) to 0.234$\pm$0.002 
(Olive et al. 1997). 
At the same time, our derived slope is significantly smaller than those of 
other authors, d$Y$/d$Z$ = 6.7$\pm$2.3 for Pagel et al. (1992) and d$Y$/d$Z$ = 
6.9$\pm$1.5 for Olive et al. This shallower slope is in good agreement 
with the value derived from stellar data for the Milky Way's disk  
 and with simple models of galactic evolution of BCGs with well-mixed 
homogeneous outflows.

We believe our $Y_{\rm p}$ value to be more reliable because we have taken 
into account several systematic effects that were not considered in 
previous work. In order of 
decreasing importance, these effects are:

1) Stellar HeI absorption lines underlying HeI emission lines. When
these are not recognized and corrected for, the derived $Y_{\rm p}$ is too low.
Underlying stellar absorption is particularly important in the BCD I Zw 
18 (compare in Figure 2 the He I line intensities in the NW component
where underlying stellar absorption is important and in the SE component
where that effect is less important). Because this BCD  
has the lowest metallicity known, it plays a particularly important role in 
determining the intercept and slope of the $Y$ versus O/H and $Y$ versus N/H 
regression lines and hence $Y_{\rm p}$.

2) To derive the He mass fraction, the emission He I line fluxes (Figure 2)
need to be corrected for several mechanisms which enhance the line 
emission.   
Previous authors usually use only the single 6678 He I line  and 
correct only for electron collisional enhancement. This  
correction is usually carried out adopting 
the electron density derived from the [S II] 6717/6731 emission-line ratio.
They neglect fluorescent enhancement which can also be important.
We use the five brightest He I lines in the optical range, which allows us 
to correct for both collisional and fluorescent enhancements and 
calculate the electron density $N_e$(He II) in a self-consistent manner.
 Setting  $N_e$(He II) equal to  
$N_e$(S II) is not physically reasonable as the S$^+$ and He$^+$ regions are 
not expected to coincide, given the large difference in the S I and He I
 ionization potentials.

3) We have observed all the galaxies in our sample with the same 
telescopes and instrumental set up,
and the data were all reduced in a homogeneous way. This is in 
contrast to more heterogeneous BCD samples used by previous investigators.
 A uniform sample is essential to minimize as 
much as possible the artificial scatter introduced by assembling different 
data sets reduced in different ways.

Instead of the statistical approach described above, we can also derive 
the primordial He abundance from accurate measurements of the He 
abundance in  a few objects selected to have very low O/H 
to minimize the amount of He manufactured in stars. 
Izotov et al. (1999) 
have carried out such a study for the two most metal-deficient BCDs known.
I Zw 18 (1/50 of solar metallicity) and SBS 0335--052 (1/41 of solar 
metallicity) provide a study in contrast concerning the 
different physical mechanisms which may modify the 
He I emission-line intensities. While in I Zw 18, the electron number density 
is small ($N_e$ $\leq$ 100 cm$^{-3}$) and collisional enhancement has a 
minor effect on the derived helium abundance, $N_e$ is much higher 
in SBS 0335--052 ($N_e$ 
$\sim$ 500 cm$^{-3}$ in the central part of the H II region). Additionally, 
the linear size of the H II region in SBS 0335--052 is $\sim$ 5 times 
larger than in I Zw 18, suggesting that it may be optically thick for 
some He I transitions. In fact, both collisional and fluorescent 
enhancements of He I emission lines play an important role in this 
galaxy. By contrast, underlying stellar He I absorption is
$\sim$ 5 times smaller in SBS 0335--052 than in I Zw 18.
Izotov et al. (1999) derive $Y$ = 0.243$\pm$0.007 for 
I Zw 18 , and $Y$ = 0.2463$\pm$0.0015 for SBS 0335--052. The 
weighted mean is then $Y$ = 0.2462$\pm$0.0015. Using d$Y$/d$Z$ = 2.4 from the 
regression lines above, the 
stellar He contribution is 0.0010, giving a primordial value $Y_{\rm p}$ = 
0.2452$\pm$0.0015, in excellent agreement with the value 0.2443$\pm$0.0015 
derived from extrapolation of the $Y$---O/H and $Y$---N/H regression lines for our 
large BCD sample. 

\section{Primordial abundances of Deuterium, Lithium and $^3$Helium}

\subsection{Deuterium}

Of all light elements, the abundance of deuterium (D) is the most sensitive 
to the baryonic mass. Stars do not make significant D but D is easily destroyed
because of its fragility, so that D/H measured 
in the interstellar medium of the Milky Way or the Solar System constitute 
only lower limits to the primordial D abundance. The latter can be 
measured directly in low-metallicity absorption line systems in the spectra 
of high-redshift quasars. The quasar is used as a background light source,
and the nearly primordial gas doing the absorbing is in the outer regions of 
intervening galaxies 
or in the intergalactic medium (the so-called Lyman $\alpha$ clouds).

Tytler and his group (see Tytler et al. 2000 for a review) have vigorously 
pursued this type of measurements. They have now obtained D/H measurements 
in the line of sight towards 4 quasars. Combining all measurements, they found
all their data are consistent with a single primordial 
value of the D/H ratio: (D/H)$_{\rm p}$ 3.0$\pm$0.4$\times$10$^{-5}$ (O'Meara et al.
2000). This latest value is about 10\% lower than their previous value 
(D/H)$_{\rm p}$ = 3.39$\pm$0.25$\times$10$^{-5}$ (Burles \& Tytler 1998). During the 
period 1994 to 1996, there were reports in the literature of D/H varying 
over a factor of 10 along different quasar line of sights.
Songaila et al. (1994) reported (D/H)$_{\rm p}$ = 24$\pm$4$\times$10$^{-5}$ in one 
system. However, later observations showed the absorption near D to be 
at the wrong velocity and being too wide. The high value is now generally 
attributed to contamination from another HI cloud with a velocity 
near that of D along 
the line of sight. Tytler et al. (2000) argue convincingly that all available 
data are consistent with a single low value of (D/H)$_{\rm p}$, and that quasar
spectra that give high (D/H)$_{\rm p}$ values can be readily interpreted as 
contamination by HI clouds along the line of the sight to the quasars and 
having a velocity near that of D.

\subsection{Lithium}  

Old halo stars that formed from nearly pristine gas with very low iron 
abundances 
during the gravitational collapse of the Milky Way show approximately 
constant $^7$Li/H (the so called ``Spite plateau'', Spite \& Spite 1982), 
implying that their $^7$Li is nearly primordial. Creation or depletion 
of $^7$Li may make the $^7$Li abundances of halo stars deviate from the 
primordial value. Creation of $^7$Li in the interstellar medium by 
cosmic ray spallation prior to the formation of the Milky Way has to be 
less than 10--20\%, so as not to produce more Be than is observed since 
the latter is also enhanced in this process (Ryan et al. 1999).

There is still considerable debate concerning the possible depletion 
of $^7$Li inside stars. Depletion mechanisms that have been proposed 
include mixing due to rotation or gravity waves, mass loss in stellar 
winds and gravitational settling.  If
depletion is present, then the primordial $^7$Li abundance is higher than the
value ($^7$Li/H)$_{\rm p}$ = (1.73$\pm$0.21)$\times$10$^{-10}$ obtained by Bonifacio 
\& Molaro (1997). Vauclair \&
Charbonnel (1998) have shown that the depletion of $^7$Li in Population II
stars may be as high as 30\%. They give for the primordial lithium
abundance ($^7$Li/H)$_{\rm p}$ = (2.24$\pm$0.57)$\times$10$^{-10}$. 
Ryan et al. (1999) also find that the small scatter in their data limits 
the mean depletion of $^7$Li to be less than 30\%. 
Pinsonneault et al. (1998) have analyzed $^7$Li depletion in rotating stars. 
They found that the depletion factor can be as high as 1.5 -- 3 times, larger 
than the value obtained by Vauclair \& Charbonnel (1998). Pinsonneault et al. 
(1999) give a primordial lithium abundance ($^7$Li/H)$_{\rm p}$ = 
(3.9$\pm$0.85)$\times$10$^{-10}$. In summary, it is likely that the 
$^7$Li abundance measured in halo stars on the Spite plateau is not 
the primordial value and needs to be corrected upwards by a factor probably 
less than 2.

\subsection{$^3$Helium}

Although the primordial abundance of $^3$He is nearly as sensitive to the 
baryon density as D, it has not been yet measured, mainly because low-mass 
stars make a lot of $^3$He, increasing its value in the interstellar medium 
of the Milky Way well above the primordial value. Furthermore, the amount of 
$^3$He destroyed in stars is unknown. Rood et al. (1998) have measured an 
average $^3$He/H = 1.6$\pm$0.5$\times$10$^{-5}$ in Galactic H II regions.
This value represents the average in the interstellar medium of the Milky Way,
but it is not known how to use it to derive the primordial abundance of 
$^3$He.    

%
%
%
\def\baselinestretch{0.5}
\begin{figure}[tbh]
\centering
\hspace*{-0.5cm}\psfig{figure=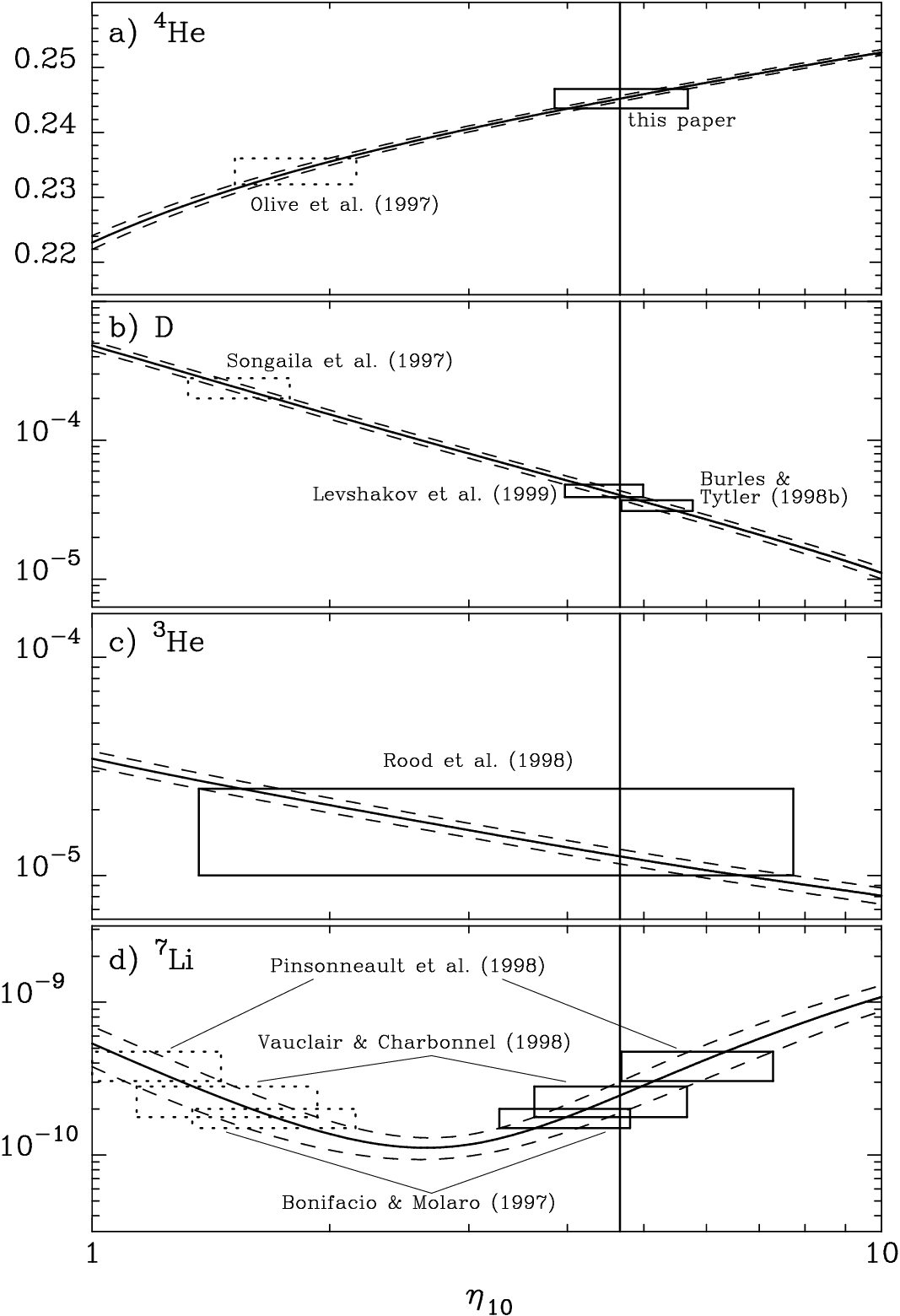,width=7.0cm,angle=0}
\hspace*{0.5cm}\psfig{figure=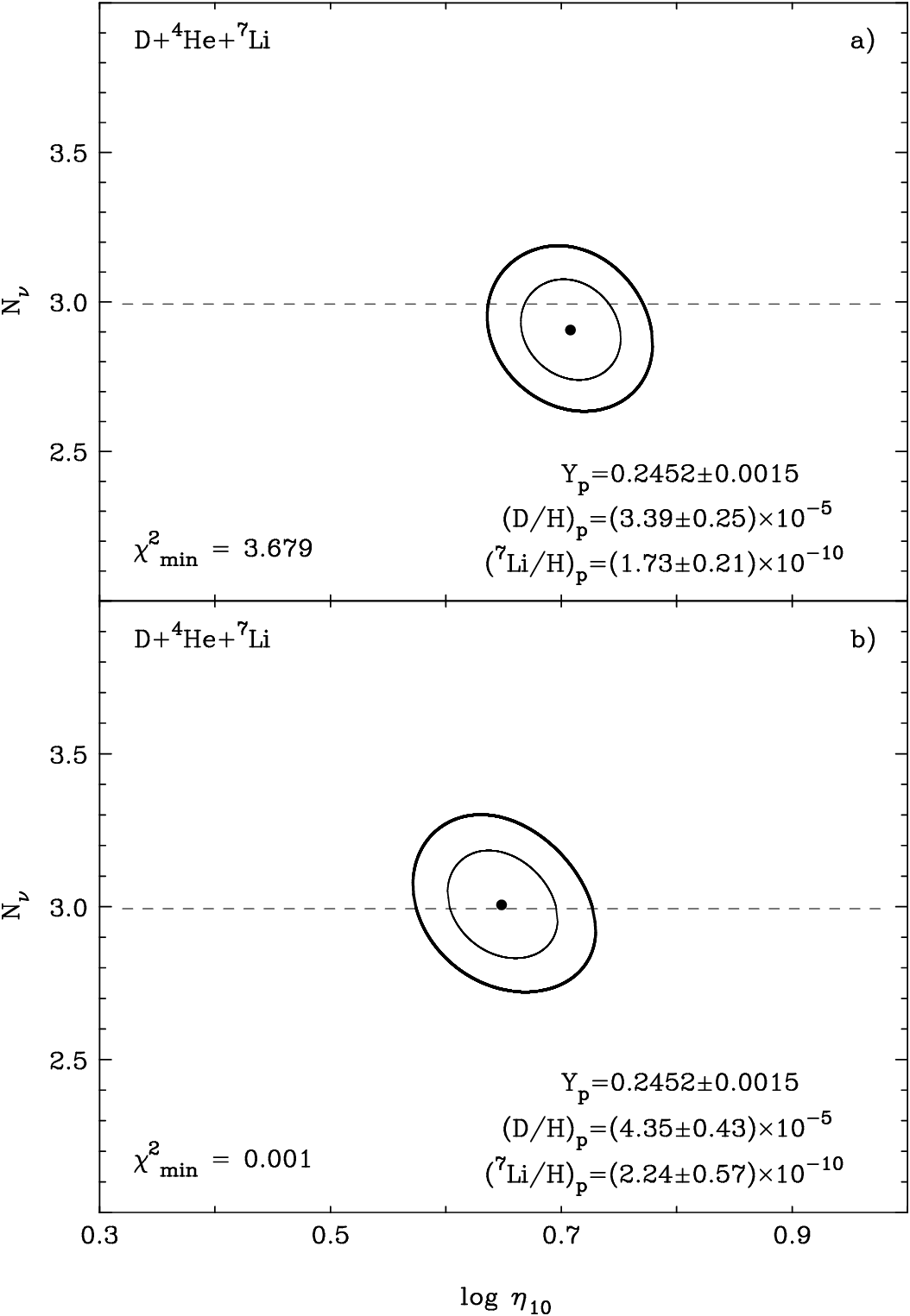,width=7.0cm,angle=0}
\caption[]{\label{Fig4}The abundance of (a) $^4$He, (b) D, (c) $^3$He 
and (d) $^7$Li
as a function of $\eta_{10}$ $\equiv$ 10$^{10}$ $\eta$, where $\eta$ is the 
baryon-to-photon 
number ratio, as given by the standard hot big bang nucleosynthesis model.
 The abundances of D, $^3$He and $^7$Li are number ratios
relative to H. For $^4$He, the mass fraction $Y$ is shown. The 
value of Izotov \& Thuan (1998) and Izotov et al. (1999) 
$Y_{\rm p}$ = 0.245$\pm$0.002 gives $\eta$ = 
(4.7$^{+1.0}_{-0.8}$)$\times$10$^{-10}$
as shown by the solid vertical line. We show other data with 1$\sigma$ boxes. }
\caption[]{\label{Fig5}Joint fits to the baryon-to-photon number ratio, 
log $\eta_{10}$, and the equivalent number of light neutrino species $N_\nu$ 
using a $\chi^2$ analysis with the code developed by Fiorentini et al. (1998)
and Lisi et al. (1999) (a) for primordial abundance values $Y_{\rm p}$ (Izotov \& Thuan 1998), 
(D/H)$_{\rm p}$ (Burles \& Tytler 1998) and ($^7$Li/H)$_{\rm p}$ (Bonifacio \& Molaro 1997) and (b) same as in (a) except (D/H)$_{\rm p}$ from Levshakov et al. (1999).
The experimental value is shown by the dashed line. }
\end{figure}
\def\baselinestretch{1.0}

\section{The consistency of Big Bang nucleosynthesis and the 
baryonic content of the Universe}

In Figure 4, 
solid lines show the primordial abundances of $^4$He, D, $^3$He and $^7$Li
predicted by standard big bang nucleosynthesis theory as a function
of the baryon-to-photon number ratio $\eta$. The dashed lines are 1$\sigma$
uncertainties in model calculations. 
The analytical fits are taken from Fiorentini et al. (1998). The solid boxes 
show the 1$\sigma$ predictions of $\eta$ as inferred from the measured primordial 
abundances of $^4$He (Izotov \& Thuan 1998, Izotov et al. 1999), 
D (Burles \& Tytler 1998), $^3$He (Rood et al. 1998) and $^7$Li
(Bonifacio \& Molaro 1997; Vauclair \& Charbonnel 1998; Pinsonneault
et al. 1998). 
For D, we have also shown the value of 
Levshakov et al. (1999). These authors have used the data of Tytler et al.
together with kinematic models with correlated turbulent 
motions for absorbing Ly$\alpha$ clouds instead of microturbulent models
to derive  (D/H)$_{\rm p}$ = (4.35$\pm$0.43)$\times$10$^{-5}$, 
slightly higher than the value of Burles \& Tytler (1998).
All these determinations are consistent to within 1$\sigma$ 
with a baryon-to-photon
number ratio $\eta$ = (4.7$^{+1.0}_{-0.8}$)$\times$10$^{-10}$, corresponding
to the primordial Helium mass fraction
$Y_{\rm p}$ = 0.245$\pm$0.002 value determined by Izotov \& Thuan (1998). 
This baryon-to-photon number ratio 
translates to a baryon mass fraction $\Omega_b$$h^2_{50}$ = 
0.068$^{+0.015}_{-0.012}$  where $h_{50}$ 
is the Hubble constant in units of 50 
km s$^{-1}$Mpc$^{-1}$.

For comparison, we have also shown with 
dotted boxes the 1$\sigma$ predictions of $\eta$ inferred from the 
``low'' primordial
abundance of $^4$He derived by Olive et al. (1997),
and the ``high' D/H value of Songaila et al. (1994),
along with the values of $^7$Li
(Bonifacio \& Molaro 1997; Vauclair \& Charbonnel 1998; Pinsonneault
et al. 1998). This second data set has, however, several problems:  
1) Olive et al. (1997) have not 
taken into account underlying stellar absorption and fluorescent enhancement,
which results in an artificially low helium mass fraction; and  
2) the high primordial D abundance reported by Songaila et al. (1994)
is probably caused by H contamination.

\section{The number of light neutrino species}

We use the statistical $\chi^2$ technique with the code described by
Fiorentini et al. (1998) and Lisi et al. (1999) to analyze the consistency of 
different sets of primordial $^4$He, D and $^7$Li abundances,
 varying both $\eta$ and the equivalent number 
of light neutrino species $N_\nu$. The lowest $\chi^2_{\rm min}$ = 0.001 
results when $\eta$ = 4.45$\times$10$^{-10}$ and $N_\nu$ = 3.0$\pm$0.3 (2$\sigma$) 
for the set of primordial abundances with $Y_{\rm p}$ = 0.245$\pm$0.002 
(Izotov \& Thuan 1998, Izotov et al. 1999), 
(D/H)$_{\rm p}$ = 
(4.35$\pm$0.43)$\times$10$^{-5}$ (Levshakov et al. 1999) and ($^7$Li/H)$_{\rm p}$ = 
(2.24$\pm$0.57)$\times$10$^{-10}$ (Vauclair \& Charbonnel 1998). 
If instead, we use the D/H of Burles \& Tytler (1998), then
$\chi^2_{min}$ = 3.7 and $N_\nu$ = 2.9$\pm$0.3.
 The joint fits of $\eta$ and $N_\nu$ to both data sets are shown in
Figure 5a and 5b, respectively. 
The 1$\sigma$ ($\chi^2$ -- $\chi^2_{min}$ = 2.3) and
2$\sigma$ ($\chi^2$ -- $\chi^2_{min}$ = 6.2) deviations are shown 
respectively by thin and thick solid lines. 
 Both values of $N_\nu$ are consistent with the experimental 
value of 2.993$\pm$0.011 (Caso et al. 1998) shown by the dashed line. 
Therefore, we conclude that both data sets of primordial abundances of light 
elements are in good agreement with predictions of standard big bang 
nucleosynthesis theory.

\acknowledgements{We are grateful for the partial financial support of 
NSF grant AST 96-16863. TXT thanks the organizers for holding such a wonderful 
meeting in his lovely native city. }

\vfill

\begin{moriondbib}

\bibitem{bm97} Bonifacio, P. \& Molaro, P. 1997, MNRAS, 285, 847
\bibitem{bt98} Burles, S., \& Tytler, D. 1998, ApJ, 507, 732
\bibitem{c98} Caso, C. et al. (Particle Data Group). 1998, Eur.J.Phys.C, 3,1
\bibitem{f98} Fiorentini, G., Lisi, E., Sarkar, S., \& Villante, F. L.
1998, Phys. Rev. D, 58, 063506 
\bibitem{it98} Izotov, Y.I. \& Thuan, T.X. 1998, ApJ, 497, 227 
\bibitem{it99} Izotov, Y.I. \& Thuan, T.X. 1999, ApJ, 511, 639
\bibitem{itl94} Izotov, Y. I., Thuan, T. X., \& Lipovetsky, V. A. 1994, ApJ, 
435, 647 
\bibitem{itl97} Izotov, Y. I., Thuan, T. X., \& Lipovetsky, V. A. 1997, ApJS, 108, 1
\bibitem{i99} Izotov, Y. I., Chaffee, F. H., Foltz, C. B.,
Green, R. F., Guseva, N. G., Thuan, T. X. 1999, ApJ, 527, 757
\bibitem{kt90} Kolb, E.W. \& Turner, M.S. 1990, The Early Universe,
(New York: Addison-Wesley)
\bibitem{l98} Levshakov, S. A., Kegel, W. H., \& Takahara, F. 1998,
MNRAS, 302, 707
\bibitem{l99} Lisi, E., Sarkar, S., \& Villante, F.L. 1999, Phys. Rev. D, 59, 123520
\bibitem{o97} Olive, K. A., Skillman, E. D., \& Steigman, G. 1997,
ApJ, 483, 788 
\bibitem{o00} O'Meara, J.M., Tytler, D., Kirkman, D., Suzuki, N., Prochaska, J.X.,
Lubin, D. \& Wolfe, A.M. 2000, astro-ph 0011179
\bibitem{p86} Pagel, B. E. J., Terlevich, R. J., \& Melnick, J. 
1986, PASP, 98, 1005
\bibitem{p92} Pagel, B. E. J., Simonson, E. A., Terlevich, R. J., \&
Edmunds, M. G. 1992, MNRAS, 255, 325
\bibitem{p74} Peimbert, M., \& Torres-Peimbert, S. 1974, ApJ, 193, 327
\bibitem{p99} Pinsonneault, M. H., Walker, T. P., Steigman, G., \&
Naranyanan, V. K. 1999, ApJ, 527, 180
\bibitem{r98}   Rood, R. T., Bania, T. M., Balser, D. S., \& Wilson, T. L.
1998, Space Sci. Rev., 84, 185
\bibitem{r99} Ryan, S.G., Norris, J.E. \& Beers, T.C. 1999, ApJ, 523, 654
\bibitem{s67} Sakharov, A.D. 1967, JETP Letters, {91 B}, 24
\bibitem{s94} Songaila, A., Cowie, L. L., Hogan, C.J., \& Rugers, M. 1994, 
Nature, 368, 599
\bibitem{s82} Spite, F. \& Spite, M. 1982, A\&A, 115, 357 
\bibitem{ti00} Thuan, T.X. \& Izotov, Y.I. 2000, in IAU Symp. 198, 
The light elements and their evolution, eds. L. da Silva, M. Spite \& J.R. de 
Medeiros (San Francisco: ASP), 176 
\bibitem{t00} Tytler, D., O'Meara, J.M., Suzuki, N. \& Lubin, D. 2000,
Physica Scripta, 85, 12
\bibitem{v98} Vauclair, S., \& Charbonnel, C. 1998, ApJ, 502, 372
 
\end{moriondbib}
\end{document}